\documentclass[aps,pra,reprint,superscriptaddress]{revtex4-1}

\usepackage{graphics} 
\RequirePackage{graphicx}
\usepackage{multirow}

\begin{document}

\title{Wavefunctions from Energies: Applications in simple potentials.}

\author{Dar\'{\i}o M. Mitnik}
\affiliation{Instituto de Astronom\'{\i}a y F\'{\i}sica del Espacio, 
CONICET--UBA, Buenos Aires, Argentina }
\author{Santiago A.H. Mitnik}   
\affiliation{Escuela de Pol\'{\i}tica y Gobierno, 
UNSAM, Buenos Aires, Argentina }
\noaffiliation

\date{\today}

\begin{abstract}
A remarkable mathematical property -- somehow hidden and recently 
rediscovered --, allows obtaining the eigenvectors of a Hermitian 
matrix directly from their eigenvalues. 
That opens the possibility to get  
the wavefunctions from the spectrum, an elusive goal of many 
fields in physics. 
Here, the formula is assessed for simple potentials, recovering 
the theoretical wavefunctions within machine accuracy. 
A striking feature of this eigenvalue--eigenvector relation  
is that it does not require knowing any of the entries of the 
working matrix. However, it requires the knowledge of the 
eigenvalues of the minor matrices (in which a row and a column have 
been deleted from the original matrix). 
We found a pattern in these sub--matrices spectra, allowing to 
get the eigenvectors analytically.
The physical information hidden behind this pattern is analyzed.
\end{abstract}


\maketitle

\section{Introduction}
\label{sec:intro} 

Recently \cite{DentonRosetta:19}, precise perturbative calculations of 
neutrino oscillation probabilities yield a surprising result: 
given the eigenvalues of the Hamiltonian and its minor sub--matrices, 
it is possible to write down all the 
probabilities by a quite simple expression.
In a following two--pages paper the authors \cite{Denton:19} described 
and proved this eigenvector--eigenvalue identity (\ref{eq:eigeig}), 
which is the subject of the present work.
Despite its simplicity and the possible implications, the identity 
was not broadly well known. 
Lately, it achieves a blaze of publicity, after a popular 
article \cite{wolcholver:19} reported the unexpected relationship 
between the eigenvalues and eigenvectors in the neutrino oscillation 
calculations.
The authors of \cite{Denton:19} rewrote the paper, adding thirty pages 
with several proofs of the identity, together with 
a discussion about its complicated history in the literature, and 
the reasons for its unawareness. 
They also showed that the equation appeared in different references, 
being independently rediscovered in diverse fields. 
In most cases, the papers were weakly connected, lacking in significant 
propagation of citations.

The {\it eigenvector--eigenvalue identity} \cite{Denton:19} 
relates the eigenvectors $v_i$ of an $n \times n$ Hermitian matrix 
$H$, with its eigenvalues $\lambda_1, \lambda_2, \ldots, \lambda_n$. 
In this work, it is slightly modified as follows:
\begin{eqnarray}
|v_{j,i} |^2 \, \, \prod_{k=1; k \neq i}^n \, 
\left( \lambda_i - \lambda_k \right) = 
\prod_{k=1}^{n-1} \, \left( \lambda_i - \lambda_k^{(j)} \right) \, ,
\label{eq:eigeig}
\end{eqnarray}
where $v_{ji}$ is the $j^{\mathrm th}$ component of the 
normalized eigenvector $v_i$ associated to the eigenvalue $\lambda_i$. 
The eigenvalues 
$\lambda_1^{(j)}, \lambda_2^{(j)}, \cdots , \lambda_{n-1}^{(j)}$ 
correspond to the minor $M_j$ of $H$, formed by removing the 
$j^{\mathrm th}$ row and column.

Identity (\ref{eq:eigeig}) has many remarkable aspects 
to discuss. 
We can mention many positive and very promising issues.
First, notice that at no point one
ever actually needs to know any of the entries of the matrix to 
calculate its eigenvectors. 
From the physical point of view, this can be very useful in systems 
in which the eigenvalues are got by measurements. 
The most direct application could be the recovering of the electronic 
densities from the experimental spectra, which is, perhaps, 
the holy grail of the physical chemistry.
Second, it can lead to the development of new numerical methods to 
diagonalize large matrices, faster and more efficiently. 
The standard computer routines use much more memory and resources 
if the eigenvectors are needed, in place of solely the eigenvalues. 

However, these two advantages have some weaknesses.
Indeed, the values of the original matrix $H$ are not needed, 
but since the expression (\ref{eq:eigeig}) involves the minors $M_j$, 
this statement is not completely straight. 
It will be undisputed, only if the minor's 
eigenvalues $\lambda_i^{(j)}$ are available, without the 
explicit knowledge of the matrix elements.
From the numerical point of view, the new algorithms do not 
seem to be competitive, unless for very large matrices.
The reason is that for every eigenvector $v_i$, all the 
eigenvalues $\lambda_k^{(j)}$ of each minor $M_j$, are needed.
Thus, the algorithm requires to obtain all the eigenvalues of 
$n-1$ different matrices, which will result in a much more 
costly way to solve an eigenproblem.

In the present work, all these aspects are examined. 
First, we prove the eigenvector--eigenvalue identity for 
the solution of the corresponding Schr\"odinger equation 
for simple one--dimensional potentials. The wavefunctions obtained 
by direct diagonalization, and by using the expression (\ref{eq:eigeig}) 
are compared for harmonic oscillators, Coulombic potentials, and 
the infinite potential well. 
Then, the spectra of the minor matrices are analyzed. 
We found that these eigenvalues are not randomly scattered, else 
they behave systematically, forming a pattern that could be 
described analytically.
Finally, a deep analysis of the minor eigenvalues $\lambda^{(j)}$ for 
the simplest potential (the infinite potential well), 
lead us to understand the origin of this singular behavior, 
which could be useful for 
the calculation of eigenvectors of other potentials.

\section{Assesment of the eigenvalue--eigenvector identity}
\label{sec:assesment} 

To corroborate the expression (\ref{eq:eigeig}) in physical 
problems, we will validate it first, in simple 
potential problems, solving the one--dimensional Schr\"odinger equation
\begin{equation}
\hat{H}(x) \, v_i(x) = \lambda_i \, v_i(x) \, ,
\end{equation}
where
\begin{equation}
\hat{H} = -\frac{1}{2}\frac{d^2}{dx^2} + V(x) \, .
\end{equation}
We choose the most commonly studied simple potentials, say, the 
harmonic oscillator
\begin{equation}
V(x) = V_{\mathrm{HO}}(x) \equiv \frac{1}{2} \omega^2 \, (x - x_0) \, ,
\label{eq:Vosc}
\end{equation}
the Coulomb potential
\begin{equation}
V(x) = V_{\mathrm{C}}(x) \equiv -\frac{Z}{x}  \, ,
\label{eq:Vcoul}
\end{equation}
and the infinite potential well
\begin{equation}
V(x)  = V_{\mathrm{W}}(x) \equiv
\left\{\begin{array}{ll}
        V_0, \,\,\, & \mathrm{for} \,\,\, x \leq L\\
        \infty,  & \mathrm{for} \,\,\, x>L
\end{array}  \right. \, .
\label{eq:Vpozo}
\end{equation}
 
For each potential, the corresponding Schr\"odinger equation was 
approximated in first--order finite differences, 
resulting in tridiagonal matrices.
First, the full Hamiltonians have been diagonalized, by using 
standard computational packages \cite{lapack}, getting the 
eigenvectors $v$ and eigenvalues $\lambda$. 
Next, we successively construct the minor matrices $M_j$, 
deleting from the full Hamiltonian matrix the $j^{\mathrm{th}}$ 
row and column. In that way, the corresponding $\lambda^{(j)}$ 
eigenvalues have been obtained.
Finally, Eq.~(\ref{eq:eigeig}) is used to get the 
``reconstructed" eigenvalues. 
The operations involved in this rebuilding require some care. 
The order of the operations is important, and also, it is better to 
add logarithms than to multiply terms having different orders of 
magnitudes.
One of the principal numerical advantages of the proposed 
approach resides in the fact that the matrices are not needed, 
else, only their eigenvalues. Therefore, we also take care to 
exploit that, performing the calculation in a particular 
sequence, in which only one minor eigenvalues are stored in 
memory, at any step. 

For all these potentials, and for many different Hamiltonian sizes, 
the reconstructed results agree with the original eigenvectors within 
machine precision. 
Thus, the first conclusion reached in the present work is that 
for any number $n$ of points used in the numerical grid, 
the eigenvalue--eigenvector identity can be considered numerically exact.

\begin{figure}[h!]
\includegraphics[width=0.45\textwidth]{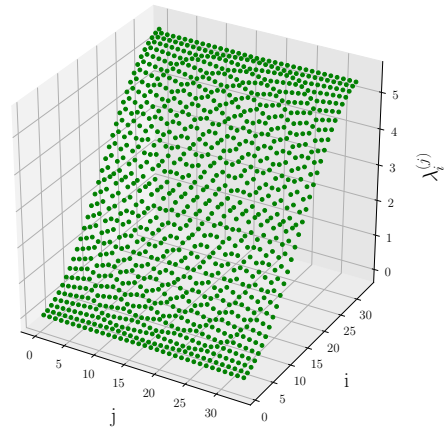} 
\caption{\label{fig:toteig}
Eigenvalues $\lambda_i^{(j)}$ of the minor matrices $M_j$, 
for the $V_{\mathrm{W}}$--Hamiltonian corresponding to 
the infinite potential well. }
\end{figure}

Of course, there is a drawback in the expression (\ref{eq:eigeig}) 
and it is the requirement of the calculation of the 
eigenvalues for $n$ different matrices, which, in general, takes 
more computational efforts than the direct diagonalization of an 
$n \times n$ matrix.
The algorithm can be improved significantly if one can found 
some relationship between the original eigenvalues $\lambda_i$ 
and the corresponding minor eigenvalues $\lambda_i^{(j)}$. 
To this end, we first plot these eigenvalues, to see 
if they follow some identifiable behavior. 
We show, in Figure~\ref{fig:toteig}, the $\lambda_i^{(j)}$ 
values corresponding to an infinite potential well 
$V_{\mathrm{W}}(x)$ (Eq.~(\ref{eq:Vpozo})) of size $L=20$ a.u. 
and amplitude $V_0=0$, 
calculated with a numerical grid having $n=33$ points. 
From this figure, we can not get any useful information, because of the 
similarities among the $\lambda_i^{(j)}$ eigenvalues, for any minor 
matrix $j$. 

\begin{widetext}
\begin{figure*}
\includegraphics[width=0.9\textwidth]{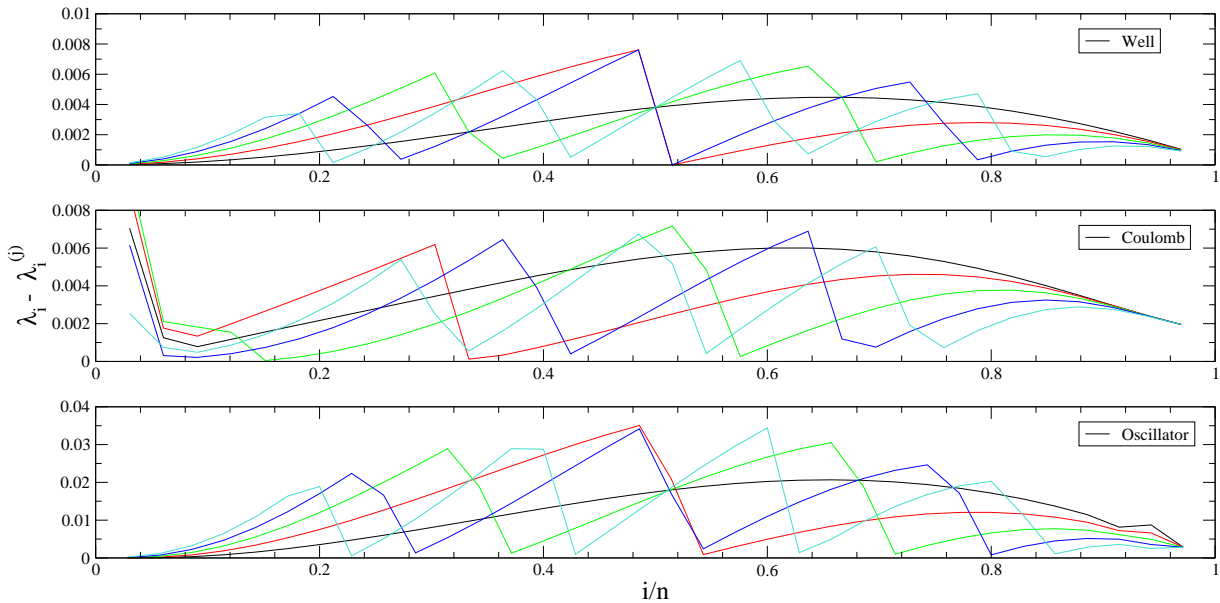} 
\caption{\label{fig:compdiff}
Differences between the full Hamiltonian eigenvalues $\lambda_i$ 
and the eigenvalues $\lambda_i^{(j)}$ of the minor matrices $M_j$
for the infinite potential well $V_{\mathrm{W}}(x)$, the 
Coulomb potential $V_{\mathrm{C}}(x)$, and the 
harmonic oscillator potential $V_{\mathrm{HO}}(x)$. \\}
\end{figure*}

\end{widetext}

For a better representation, we define and plot, 
in Figure~\ref{fig:compdiff} the differences 
$D(i,j) \equiv \lambda_i-\lambda_i^{(j)}$, 
which could be helpful to understand 
the meaning of the minor's eigenvalues. 
These differences have been calculated for three different potentials, 
an infinite potential well, a Coulomb potential, and a harmonic 
oscillator potential. 
An unexpected regular pattern appears in the three cases, 
encouraging the search for analytical expressions capable to 
represent these values.

\section{Analytical expression for the minor's eigenvalues}
\label{sec:analytical} 

Although the curves shown in Fig.~\ref{fig:compdiff} suggest the 
possibility to find analytical functions to represent the minor's 
eigenvalues, this task is not straightforward. 
Let us analyze the eigenvalues of the minor matrices for the 
infinite potential well $V_{\mathrm{W}}(x)$, which seems to be 
the potential having the most regular behavior. 
It is important to stress that we are dealing with the numerical 
solutions of a first order finite--differences approximation, 
which are different than the exact solutions. We will first 
attempt to understand the state of affairs in the numerical solutions, 
and will treat the exact solutions in further works (see a brief 
discussion in Appendix~\ref{app:analytic}).

\begin{figure}
\includegraphics[width=0.45\textwidth]{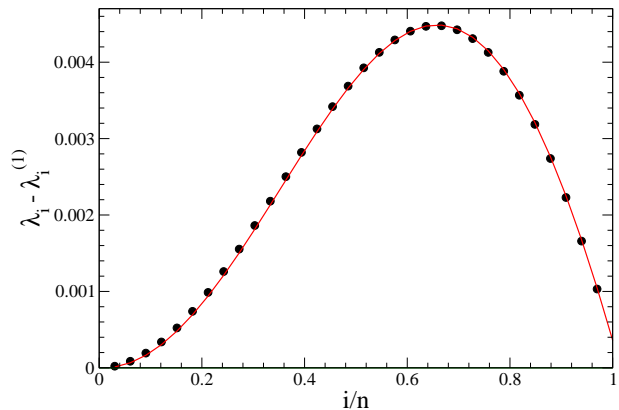} 
\caption{\label{fig:test1}
Differences $\lambda_i - \lambda_i^{(1)}$ for the minor matrix $M_1$ 
for the infinite potential well $V_{\mathrm{W}}$--Hamiltonian.
Points: Values obtained by direct diagonalization. Solid curve: 
analytic approximation. }
\end{figure}

For the first minor matrix $M_1$, the differences $D(i,1)$ 
are arranged in a smooth curve which is easily approximated, 
as is shown in Figure \ref{fig:test1}, by the expression
\begin{equation}
D(i,1) \equiv \lambda_i-\lambda_i^{(1)} \approx A_1 \, i \,
\sin\left(\frac{\pi \, i}{a_1}\right) \, ,
\label{eq:xsinx}
\end{equation}
where $a_1$ is very close to $n$, the 
number of points in the numerical grid.

The analytic approximation is more difficult for the next 
minor matrix $M_2$. 
Here, there are two different noticeable regions $d$, 
and each one should be approximated separately, i.e.,
\begin{equation}
D^{(d)}(i,j) \equiv \lambda_i-\lambda_i^{(j)} \,\, |_{i \in d} 
\approx A_j^d \, i \, 
\sin\left(\frac{\pi \, i}{a_j^d}\right) \, .
\label{eq:xsinxd}
\end{equation}

\begin{figure}
\includegraphics[width=0.45\textwidth]{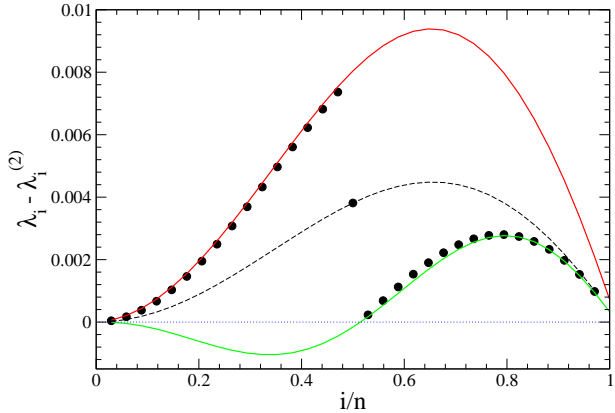} 
\caption{\label{fig:test2-34}
Differences $\lambda_i - \lambda_i^{(2)}$ for the minor matrix $M_2$, 
for a $V_{\mathrm{W}}$--Hamiltonian matrix with $n=34$ points.
Points: Values obtained by direct diagonalization. Solid curve: 
analytic approximation. Dashed curve: differences 
$\lambda_i - \lambda_i^{(1)}$. }
\end{figure}

For $j=2$ the first coefficient $a_2^1 = a_1 \approx n$, but, for the 
second half of the eigenvalues ($d=2$), the appropriate value is 
$a_2^2 \approx \frac{n}{2}$ 
(formal expressions for $a_j^d$ are given in  
Appendix \ref{app:formulas}). 
\begin{figure}
\includegraphics[width=0.45\textwidth]{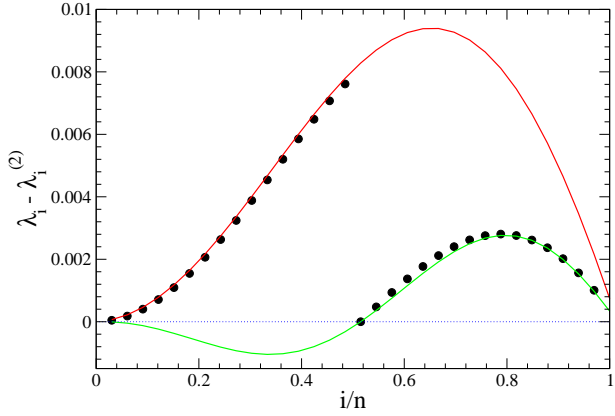} 
\caption{\label{fig:test2-33}
Differences $\lambda_i - \lambda_i^{(2)}$ for the minor matrix $M_2$.
Points: Values obtained by direct diagonalization. Solid curve: 
analytic approximation. 
The size of the $V_{\mathrm{W}}$--Hamiltonian matrix is $n=33$. 
\\ }
\end{figure}
The complications related to pursuing an understanding of the 
behavior of the minor's eigenvalues do not end here. 
We found that the size of the Hamiltonian also affects how the 
$D(i,2)$ are approximated.
Let see these values for a case in which the number of points in 
the numerical grid is even. 
Figure \ref{fig:test2-34} shows the differences between the eigenvalues 
of a Hamiltonian matrix with $n=34$ points (in place of the 
previous values showed in Fig.~\ref{fig:test2-33}, where $n=33$).
Here, the differences for the first 16 eigenvalues $D^{(1)}(i,2)$ are 
approximated with (\ref{eq:xsinxd}) using a set of parameters 
$A_j^d$ and $a_j^d$, 
and the last 16 $D^{(2)}(i,2)$ by using another set, 
like in the previous case. 
But, the difference $D(17,2) = \lambda_{17} - \lambda_{17}^{(2)}$ 
does not belong to either of both curves. 
Surprisingly, we found that this value agrees exactly 
with the difference got for the first minor, 
which is $D(17,1)=\lambda_{17} - \lambda_{17}^{(1)}$.

\begin{figure}
\includegraphics[width=0.45\textwidth]{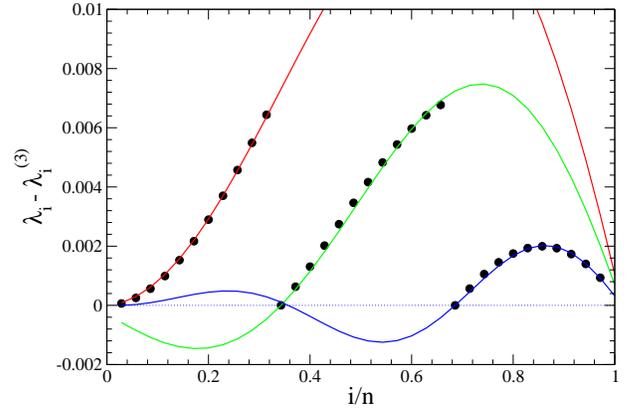} 
\caption{\label{fig:test3-35}
Differences $\lambda_i - \lambda_i^{(3)}$ for the minor matrix $M_3$, 
for a $V_{\mathrm{W}}$--Hamiltonian matrix with $n=35$ points. \\
\vspace{0.01\textheight} }
\end{figure}

\begin{figure}
\includegraphics[width=0.45\textwidth]{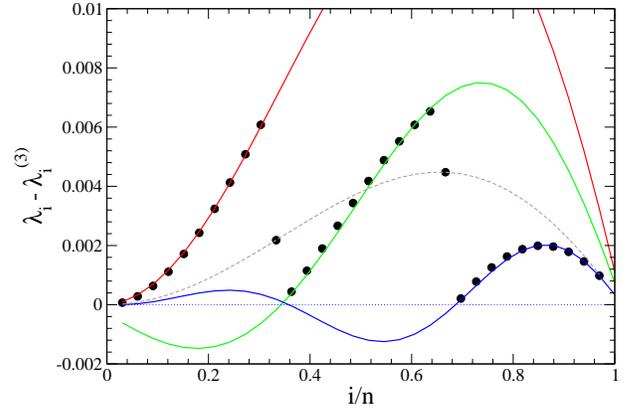} 
\caption{\label{fig:test3-33}
Differences $\lambda_i - \lambda_i^{(3)}$ for the minor matrix $M_3$, 
for a $V_{\mathrm{W}}$--Hamiltonian matrix with $n=33$ points. 
Dashed curve: $\lambda_i - \lambda_i^{(1)}$.  }
\end{figure}

\begin{figure}
\includegraphics[width=0.45\textwidth]{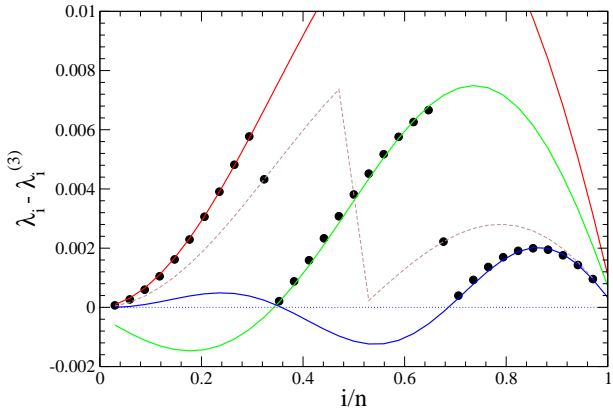} 
\caption{\label{fig:test3-34}
Differences $\lambda_i - \lambda_i^{(3)}$ for the minor matrix $M_3$, 
for a $V_{\mathrm{W}}$--Hamiltonian matrix with $n=34$ points. 
Dashed curve: $\lambda_i - \lambda_i^{(2)}$.  }
\end{figure}

More intricate is the approximation for the $M_3$ minor's 
eigenvalues ($j=3$).
We start the analysis with a Hamiltonian matrix having $n=35$ points.
As is shown in Figure \ref{fig:test3-35}, we need to 
establish three different regions $d$, where the size of each one is 
pretty near $1/3$ (see Appendix \ref{app:formulas}).  
Since 34 is not divisible by 3, we do not know beforehand how 
many points $i$ belong to each region $d$.
We found that the approximation (\ref{eq:xsinxd}) 
works well for the three ranges if considering 11 points 
belonging to the first and last regions and 12 points to the middle.
The corresponding parameters $a$ are roughly $a_3^1 \approx n$, 
$a_3^2 \approx \frac{2n}{3}$, $a_3^3 \approx \frac{n}{3}$.
As pointed out before, this distribution is particular for 
a given Hamiltonian size.
The $n=33$--case is illustrated in 
Figure \ref{fig:test3-33}.
Here, 10 eigenvectors can be fitted very well with (\ref{eq:xsinxd}), 
for each region $d$.
We run into troubles at two points corresponding 
to $i=\frac{n}{3}$ and $i=\frac{2n}{3}$.
As shown in the figure, the same unexpected property found before  
holds for this case: for these points, the eigenvalue differences 
for $M_3$ coincide exactly with the corresponding values of the 
first minor $M_1$.
Nevertheless, the picture is different for $n=34$, 
as is shown in Figure \ref{fig:test3-34}. 
Here, the first and last region holds 
10 points, the central region 11 points, and, again, 
two values lie outside these curves. 
Now, these values agree exactly with those corresponding to $M_2$.

In conclusion, the eigenvalue differences $D^{(d)}(i,j)$ 
follow the general behavior given in  Eq.(\ref{eq:xsinxd}), 
in which the width coefficients 
\begin{equation}
a_j^d \approx \frac{n}{j} \left( j - d + 1 \right) \, .
\label{eq:ajd}
\end{equation}
But, it seems 
very difficult to assign a simple formula for each minor index $j$, 
for each region $d$, and for any number of points $n$.

We emphasize that the expression (\ref{eq:ajd}) is approximate. 
An accurate fitting for the $a$ values can be done, requiring 
a lot of work since adjustments must be done  
to generalize it for every number of points $n$.
These corrections are small and will not be significant for 
large matrices.
However, establishing the number of points assigned to each sector $d$ 
is not straightforward since it depends on the number of points $n$ 
and varies for each minor index $j$. 
Assigning a wrong number of points to a particular region accumulates 
enormous errors in the reconstruction formula (\ref{eq:eigeig}), 
turning any approximation of $D$ useless.
Furthermore, there are many problematic points $k$, that do not belong 
to any of the regions $d$. The $D(k,j)$ differences for these points 
agree with the values calculated for other minors $D(k,p)$, where 
the minor index $p$ also depends on $j$ and the number of points $n$. 
We found the analytical expression that relates these indexes:
\begin{equation}
p_n^{(j)} = \mathrm{mod}(n+1-j,j)
\label{eq:pmod}
\end{equation}

\begin{figure}
\includegraphics[width=0.45\textwidth]{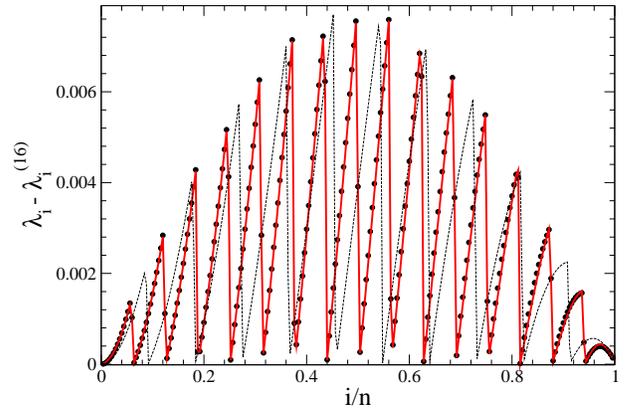} 
\caption{\label{fig:test16-250}
Differences $\lambda_i - \lambda_i^{(16)}$ for the minor matrix $M_{16}$, 
for a $V_{\mathrm{W}}$--Hamiltonian matrix with $n=250$ points. 
Dashed curve: $\lambda_i - \lambda_i^{(11)}$.  }
\end{figure}

Taking all these aspects into consideration, we can 
reproduce all the $D$ values, and thus, all the eigenvalues 
$\lambda_i^{(j)}$ belonging to all the minor matrices $M_j$, 
for any Hamiltonian size $n$. 
As an example, we illustrate, in Figure \ref{fig:test16-250}, 
the eigenvalue differences 
$D(i,16)=\lambda_i - \lambda_i^{(16)}$ for the minor matrix $M_{16}$, 
for a Hamiltonian matrix with $n=250$ points. 
All the points $k$ that are not fitted with the curves represented in 
solid lines have exactly the values $D(k,11)$ 
corresponding to the minor matrix $M_{11}$ (dashed curves), 
in agreement with our findings, since $p=\mathrm{mod}(235,16)=11$.

\section{Physical interpretation of the minor's eigenvalues}
\label{sec:physical} 

Having decoded the startling pattern conformed by the 
minor's eigenvalues, many questions remain open. 
The behavior discovered is particular to the infinite potential well. 
From Figure~\ref{fig:compdiff} it is reasonable to conjecture 
about the existence of similar relationships on the other potentials.
All the eigenvalues have been calculated numerically. 
A question arises is whether the same pattern holds also for 
the exact analytical results.
The most important issue to solve is the possibility to recognize 
a physical meaning in our findings. Perhaps, understanding 
this matter could help to answer the other questions.

The first physical realization of the eigenvector--eigenvalue identity 
has been done by Voss and Balloon \cite{Voss:20}. 
They showed that one--dimensional arrays of coupled resonators, 
described by square matrices with real eigenvalues, provide 
simple physical systems where this formula can be applied in practice.
The subsystems consist of arrays with the $j^{\mathrm{th}}$ 
resonator removed. 
Thus, from their spectra alone, the oscillation modes of the 
full system can be obtained.

\begin{figure}
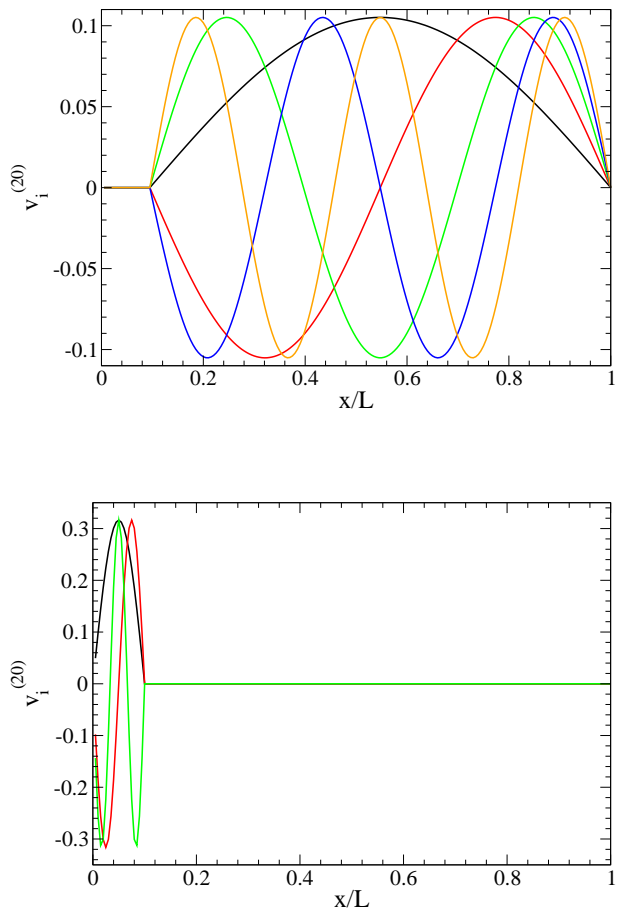

\includegraphics[width=0.45\textwidth]{v20.eps} \\
\vspace{0.05\textheight}
\includegraphics[width=0.45\textwidth]{v20z.eps} 
\caption{\label{fig:pozoM20}
Upper part: first eigenvectors $v_i^{(20)}$ for the minor 
matrix $M_{20}$, for a $V_{\mathrm{W}}$--Hamiltonian matrix 
with $n=200$ points.
Lower part: eigenvectors $v_{10}^{(20)}$, $v_{20}^{(20)}$, and 
$v_{30}^{(20)}$. }
\end{figure}

Concerning the infinite potential well, the physical meaning of 
the removal of the first row and column is very simple: 
it conforms another infinite potential well whose width decreases 
from $L$ to $L-\Delta x$, where $\Delta x$ 
is the numerical mesh size. The same potential may be thought of as 
an infinite well of size $L$, with an additional infinite wall 
located at $\Delta x$.
Accordingly, the removal of the $j^{{\mathrm th}}$ row and column 
means an infinite potential well with an infinite wall at $x=j \Delta x$.
As an example, we plot the first eigenvectors of the submatrix 
$M_{20}$, for an infinite potential well represented by $n=200$ points, 
in Figure~\ref{fig:pozoM20} (above). 
In the relative coordinates it means that an infinite wall 
is present at $x/L=0.1$. 
As is shown in the figure the first eigenvectors correspond to 
the eigenvectors of a $L - L/10$ infinite potential well. 
We also have to consider the other potential well of size 
$L/10$, which, for this particular case have eigenvalues 
coincident with the full Hamiltonian spectra for every $i=10$ functions. 
The lower part of Figure~\ref{fig:pozoM20} shows 
the eigenvalues $v_{10}^{(20)}$, $v_{20}^{(20)}$, and 
$v_{30}^{(20)}$, having the same energy as  $v_{10}$, $v_{20}$, and 
$v_{30}$ from the full Hamiltonian.

\section{Conclusions}

The eigenvalue--eigenvector identity expresses a curious and 
surprising relationship between the eigenvector of a 
matrix and the eigenvalues of its minors. 
It can be very useful from the numerical point of view, 
encouraging the study of new techniques to solve large eigenproblems. 
In general, it is much simple and fast to calculate the 
eigenvalues than the eigenvectors. Moreover, for very large 
matrices, the memory storage requirement for standard eigenvectors 
solver can lead to serious problems, which could be overcome by 
this approach.
A critical element to strengthening the use of the identity 
is the ability to avoid the calculation of the eigenvalues 
of all the minor matrices. 

In this work, we first tested the eigenvalue--eigenvector 
identity on different systems.
The formula has been used to 
reconstruct, numerically, the eigenvectors of simple 
one--dimensional Hamiltonians. 
Using only the eigenvalues of the minor matrices, the 
formula reproduces the eigenvectors within machine 
precision, even for large cases.
In most cases, the eigenvalue $\lambda_i$ is close 
to the $\lambda_i^{(j)}$ value. However, analyzing 
their differences $D(i,j)=\lambda_i - \lambda_i^{(j)}$ , 
we found systematic patterns, allowing 
to extract analytically the minor's eigenvectors without 
diagonalizing all the submatrices.

To understand the patterns, we focused 
our study on the infinite potential well. 
Elucidating the behavior of the differences $D(i,j)$ is 
not straightforward since they depend on the eigenvalue 
index $i$, the minor index $j$, and also on the number $n$ of 
points in the numerical grid representing the Hamiltonian.
We found that, indeed, it is possible to find an analytical 
expression for $D(i,j)$, and therefore, for all the 
$\lambda_i^{(j)}$.
The existence of regular patterns suggest that 
some hidden information about the eigenvectors of the 
full Hamiltonian is buried 
under the minor matrices' eigenvalues structure.
A crucial aspect to understand this information is to 
found the physical reasons that generate those patterns.
For the case studied here, the  minor matrix represents a 
clear physical case that consists of the same Hamiltonian 
but with an infinite wall located at the point corresponding 
to the index $j$.
The generalization to other potentials in 3 dimensions and 
for real scenarios, can be a matter of further interesting development.

\begin{acknowledgments}
This work was supported with PIP N$^{\circ}11220130100607$ of CONICET,
Argentina.
\end{acknowledgments}

\vspace{12pt}
The data that support the findings of this study are available from 
the corresponding author upon reasonable request.
\vspace{12pt}


\appendix
\section{Analytical expression for the approximation formulas}
\label{app:formulas}

In first--order approximation, the numerical eigenvalues 
$\lambda_i$ of the one dimension infinite potential well, are related 
to the eigenvalues $\lambda_i^{(j)}$ of the minor submatrix 
$M_j$ in the following way.
Let us define a variable 
\begin{equation}
x \equiv \frac{i}{n} \, , 
\label{eq:x}
\end{equation}
where $i$ is the 
eigenvalue index and $n$ is the number of grid points 
representing the eigenvectors.
Numerically, an infinite potential well represented by an 
$n$--points Hamiltonian has $n$ different eigenvalues, 
therefore, $x \in \{0,1\}$.
The eigenvalue differences $D(i,j)=\lambda_i-\lambda_i^{(j)}$ 
have been found to cluster in $j$ different sections, where the size 
of each of them is approximately $n/j$.
More precisely, each section begins at a point
\begin{equation}
x_{0_j}^{(d)}= \frac{d-1}{j} \left(1 + \frac{1}{n} \right) \, .
\label{eq:x0}
\end{equation}
It is worth to note that for the infinite potential well, 
the $x_{0_j}$ points correspond to the nodes of the 
$j^{\mathrm{th}}$ eigenvector.
 
The argument of the $\sin$ functions in Eq.~(\ref{eq:xsinxd})
has a wavelength proportional to a width of
\begin{equation}
a_d^{(j)} = 1 - x_{0_j}^{(d)} + \frac{1}{2 n} \, .
\end{equation}
 
The differences can be approximated by the expression
\begin{equation}
D^{(d)}(i,j) = A_j^{(d)} \, 
\sin\left( \frac{x - x_{0_j}^{(d)} }{a_d^{(j)} } \pi \right) \, (x + \Delta^{(d)}) \, .
\label{eq:Dijd}
\end{equation}
The parameter $\Delta^{(d)}$ 
\begin{equation}
\Delta^{(d)} = 
\left\{\begin{array}{ll}
        \frac{d}{2 n}, \,\,\, & \mathrm{for} \,\,\, d<\frac{j}{2}\\
        \frac{j-d+2}{2 n},  & \mathrm{for} \,\,\, 
              \frac{j}{2}\le d < j-1\\
        -\frac{1}{2 n},  & \mathrm{for} \,\,\, d=j-1
\end{array}  \right. \, ,
\end{equation}
and the coefficients $A_j^{(1)}=j$, $A_j^{(j-1)}=\frac{1}{j}$, 
and in any other case
\begin{equation}
A_j^{(d)} = \frac{j}{d} \, \,
\frac{ \sin \left( \frac{d}{j} \pi \right) }
{ \sin \left( \frac{\pi}{j-d+1} \right)  }  \, .
\end{equation}

It is very important to identify the limits of every sector $d$: 
the initial eigenvalue index $i_{0_j}^{(d)}$, the last point 
$i_{b_j}^{(d)}$, and the points $k_j^{(d)}$ that lie outside each  
region $d$. 
For simplicity, let us drop the indexes $i$ and $d$ in Eq.~(\ref{eq:x0}), 
so, every section is bounded by the range
\begin{equation}
x_0 \equiv \frac{d-1}{j} \left( 1 + \frac{1}{n} \right) < x < 
x_b \equiv \frac{d}{j} \left( 1 + \frac{1}{n} \right) \, .
\end{equation}
The corresponding $(x \times n)$ values are not necessarily integer numbers.
We define the indexes $i_{0}$ and $i_{b_z}$ as the integer part of 
$(x_0 \times n)$ and $(x_b \times n)$, respectively.
Then, we designate the following quantities:
\begin{equation}
\left( x_1,x_2,x_3 \right) \equiv 
\left( \frac{i_{b_z}-1}{n},\frac{i_{b_z}}{n},\frac{i_{b_z}+1}{n} \right) \, .
\end{equation}
The index $i_b$ is defined as 
\begin{equation}
i_b = 
\left\{\begin{array}{ll}
        i_{b_z}-1 & \mathrm{for} \,\,\, x_1 < x_b \le x_2 \\
        i_{b_z}  & \mathrm{for} \,\,\, x_2 < x_b \le x_3  
\end{array}  \right. \, ,
\end{equation}
Every sector contains the eigenvalue indexes between
\begin{equation}
i_0 + 1 \le i \le i_b - 1  \, .
\end{equation}
As explained above, the value of $D(i_b,j)^{(d)}$ is calculated 
by using Eq.~(\ref{eq:pmod}), i.e., the value corresponding to 
the index $k=i_b$ is $D(k,j)^{(d)}=D(k,j_p)^{(p)}$ where
\begin{eqnarray}
p  &=& \mathrm{mod}(n+1-j,j)   \,\,\,\, \mathrm{and}  \nonumber \\
j_p &=&  \frac{ i_b \, p  }{n} \, . 
\end{eqnarray}

\section{Implementation in the exact solutions}
\label{app:analytic}

Here, we will show how to proceed for the reconstruction of the 
exact wavefunctions (not the numerical solutions, as before). 
Again, we want to get the eigenvectors of an infinite potential well, 
using the eigenvector--eigenvalue identity.
The expression (\ref{eq:eigeig}) involves a product, so, we must  
decide, first, how many points should be used in it. 
This decision also determines which minor matrices are included, and 
their eigenvalues.
As stated above, the minor matrix $M_j$ represents the original 
Hamiltonian matrix with an infinite wall at the 
corresponding position $x_j$.
This wall separates the potential in two regions, in our case, two infinite 
wells, one from 0 to $x_j$, and the other from $x_j$ to $1$.
To reconstruct the eigenvectors through the eigenvector--eigenvalue 
identity, one must be aware to intercalate appropriately the  
energies of both wells.

As an example, let us pick 10 terms in the product (\ref{eq:eigeig}) to 
reconstruct the eigenvectors of an infinite potential well having 
a width $L=11$ a.u.. 
The walls are located at $x_j=1,2,3,\ldots,9,10$ a.u..
For instance, for the minor $M_4$, we need to calculate the spectra for two 
wells: one having a width $L_1=7$ and the other with $L_2=4$.
Their corresponding energies are 
\begin{equation}
\epsilon_{n}^{(i)} =
\frac{1}{2} \, \left( \frac{n_i \, \pi}{L_i} \right)^2 \, .
\label{eq:pozo}
\end{equation}
For the box having $L_2=7$ a.u., the lowest energies are 
$\epsilon_1^1=0.1007$, $\epsilon_2^1=0.4028$, and $\epsilon_3^1=0.9064$ a.u., 
and for $L_1=4$ a.u., the lowest energies are
$\epsilon_1^2=0.3084$, $\epsilon_2^2=1.2337$, and $\epsilon_3^3=2.7758$ a.u.. 
The ordered eigenvalues $\lambda_i^{(4)}$ are stated in 
Table~\ref{tab:eigenvalues}, together with the other values. 
The rest of the eigenvalues are got by considering the symmetry of the 
potentials, i.e., 
$\lambda_i^{(6)}=\lambda_i^{(5)}$, 
$\lambda_i^{(7)}=\lambda_i^{(4)}$, 
$\lambda_i^{(8)}=\lambda_i^{(3)}$, 
$\lambda_i^{(9)}=\lambda_i^{(2)}$, and 
$\lambda_i^{(10)}=\lambda_i^{(1)}$. 

\begin{figure}
\vspace{0.02\textheight}
\includegraphics[width=0.45\textwidth]{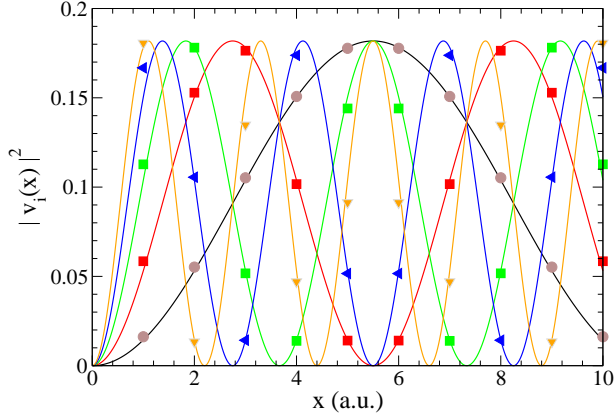} 
\caption{\label{fig:reconstr10}
First 5 exact eigenvalues $v_i(x)$ corresponding to an infinite 
potential well having a width $L=11$ a.u.. 
The curves represent $|v_i(x)|^2$. 
The points are the $|v_{ji}|^2$ values, calculated through the 
eigenvalue--eigenvector identity (\ref{eq:eigeig}) (normalized).}
\end{figure}

Using these eigenvalues, the eigenvector--eigenvalue equation reproduces 
the eigenvectors (at the positions $x_j$), without normalization.
The normalized results are shown in 
Figure~\ref{fig:reconstr10}, for the first five eigenvalues, 
demonstrating that the identity can also be used for exact analytical 
cases.

\begin{widetext}

\begin{table}[h]
\caption{Lowest eigenvalues $\lambda_i$ for the infinite potential well
having a width $L=11$ a.u., and eigenvalues $\lambda_i^{(j)}$ of 
the corresponding minor $M_j$ matrices. For each case, the 
quantum number $n_i$ and the width of the corresponding well $L_i$ 
used in Eq.~(\ref{eq:pozo}), are indicated.}
\vspace{0.01\textheight}
\begin{tabular}{c|c|c|c|c|c|c|c|c|c|} 
\cline{2-10}
  & $\lambda_1^{(j)}$  $(L_i,n_i)$ &
    $\lambda_2^{(j)}$  $(L_i,n_i)$ &
    $\lambda_3^{(j)}$  $(L_i,n_i)$ &
    $\lambda_4^{(j)}$  $(L_i,n_i)$ &
    $\lambda_5^{(j)}$  $(L_i,n_i)$ &
    $\lambda_6^{(j)}$  $(L_i,n_i)$ &
    $\lambda_7^{(j)}$  $(L_i,n_i)$ &
    $\lambda_8^{(j)}$  $(L_i,n_i)$ &
    $\lambda_9^{(j)}$  $(L_i,n_i)$ \\ \hline
\multicolumn{1}{|c|}{H} &
0.0408  (11,1) &
0.1631  (11,2) &
0.3671  (11,3) &
0.6525  (11,4) &
1.0196  (11,5) &
1.4682  (11,6) &
1.9984  (11,7) &
2.6101  (11,8) &
3.3035  (11,9) \\ \hline
\multicolumn{1}{|c|}{$M_2$} &
0.0493 (9,1)  &
0.1974 (9,2)  &
0.4441 (9,3)  &
0.7896 (9,4)  &
1.2337 (2,1)  &
1.7765 (9,5)  &
2.4181 (9,6)  &
3.1583 (9,7)  &
3.9972 (9,8)  \\ \hline
\multicolumn{1}{|c|}{$M_3$} &
0.0609 (8,1)  &
0.2437 (8,2)  &
0.5483 (3,1)  &
0.9748 (8,3)  &
1.2337 (8,4)  &
1.5231 (8,5)  &
2.1932 (3,2)  &
2.9853 (8,6)  &
3.8991 (8,7)  \\ \hline
\multicolumn{1}{|c|}{$M_4$} &
0.1007 (7,1)  &
0.3084 (4,1)  &
0.4028 (7,2)  &
0.9064 (7,3)  &
1.2337 (4,2)  &
1.6114 (7,4)  &
2.5178 (7,5)  &
2.7758 (4,3)  &
3.6256 (7,6)  \\ \hline
\multicolumn{1}{|c|}{$M_5$} &
0.1371 (6,1)  &
0.1974 (5,1)  &
0.5483 (6,2)  &
0.7896 (5,2)  &
1.2337 (6,3)  &
1.7765 (5,3)  &
2.1932 (6,4)  &
3.1583 (5,4)  &
3.4269 (6,5)  \\ \hline
\hline 
\label{tab:eigenvalues}
\end{tabular}
\end{table}
\end{widetext}


\end{document}